# Magnetic interactions as a pivotal determinant in stabilizing a novel $Ag^{II}Ag^{III}F_5$ polymorph with high-spin $Ag^{III}$

Daniel Jezierski[1*] and Wojciech Grochala[1*]

[1] *Center of New Technologies, University of Warsaw, 02089 Warsaw, Poland*
*d.jezierski@cent.uw.edu.pl, w.grochala@cent.uw.edu.pl*

*This work is dedicated to prof. Giuseppe Resnati at his 70th birthday*

**ABSTRACT**
Based on theoretical calculations, we introduce a new $Ag^{II}Ag^{III}F_5$ monoclinic polymorph with a rare high-spin $Ag^{III}$. Our analysis of the experimental x-ray diffraction data available in the literature reveals that this polymorph was likely prepared in the past in a mixture with the triclinic form of the same compound. Theoretical calculations reproduce very well the lattice parameters of both forms. Calculations suggest that under ambient conditions, the monoclinic form is the more energetically stable phase of $Ag_2F_5$. We predict a strong one-dimensional antiferromagnetic superexchange between silver cations of different valences with superexchange constant of –207 meV (hybrid functional result). The polymorph with high-spin $Ag^{III}$ owes its stability over the one with low-spin $Ag^{III}$, to these magnetic interactions.

**INTRODUCTION**
Remarkable structural and electronic similarities between $Ag^{II}$-F and $Cu^{II}$-O systems[1–4], have led to the intense exploration of fluoroargentates(II) as unconventional superconductivity (SC) precursors. Recent theoretical studies have shown how to strengthen the magnetic superexchange (SE) interactions[5,6] in these materials and allowed for estimation of the critical superconducting temperature ($T_C$) at optimum doping[7,8]. An important target here is to stabilize intermediate oxidation state of silver cations in periodic systems[9]. However, it is still unclear how could hole- or electron-doping be performed in practice. For example, compounds at formally non-integer oxidation state of silver are known, but they all feature mixed- and not intermediate valences. This family encompasses $Ag^I Ag^{II} F_3$[10], $Ag^I_2 Ag^{II} F_4$[10], $Ag^{II} Ag^{III}_2 F_8$[11] and $Ag^{II} Ag^{III} F_5$[11,12]. The latter stoichiometry is of interest for the current study.

In 1991, Žemva *et al.* first synthesized maroon $Ag_2F_5$ via a precipitation reaction at 20 °C between $AgFAsF_6$ and $KAgF_4$ in anhydrous HF, deducing its formation from a 1:1 reagent ratio and mass balance matching $Ag_2F_5$ stoichiometry[11]. However, only the relative intensities and positions of diffraction reflexes were then reported, with no further structural details provided. In 2002, Fischer and Müller accidentally prepared dark brown $Ag_2F_5$ via solvothermal reaction in HF and they have determined its crystal structure in the space group $P\bar{1}$ from single-crystal X-ray measurements[13]. Here, $Ag_2F_5$ was a byproduct of a solvothermal reaction between $Ag_2O$ and $RhCl_3$ (at 1:2 ratio and aimed at synthesizing $AgRhF_6$[13]), conducted in a sealed Monel autoclave with anhydrous HF saturated with gaseous fluorine at ~450°C for five days. The crystal structure features two distinct metal centers – $Ag^{II}$ ($d^9$, s=1/2) and a low-spin $Ag^{III}$ ($d^8$, s=0). The $Ag^{II}$ cations are linked via fluoride bridges in infinite $[AgF^+]$ chains which feature strong antiferromagnetic interactions – a prerequisite for achieving superconductivity in cuprates[14]. The low-spin $Ag^{III}$ cations are magnetically silent leading to the formulation of $Ag_2F_5$ as $[Ag^{II}F][Ag^{III}F_4]$ with square-planar $[Ag^{III}F_4^-]$ anions.

In this work we focus on a new *C*2/*c* polymorph of $Ag_2F_5$ compound with HS-$Ag^{III}$. This species is analogous and isostructural to the previously described $Ag^{II}Co^{III}F_5$[15] but trivalent cobalt cation is substituted by silver. Our theoretical study is based on quantum mechanical calculations using DFT (density functional theory) approach, and it focuses on the similarities and differences in the crystal, phonon, electronic and magnetic structures of *C*2/*c* and $P\bar{1}$ forms and their relative stability. Furthermore, based on analysis of available x-ray diffraction data, we claim that Žemva et al.[11] have obtained a mixture of both polymorphs of $Ag_2F_5$ in their experiments rather than a single phase.

**METHODS**
All calculations were carried out within within the Kohn–Sham DFT framework using VASP 5.4.4[16] We employed the PBEsol form of the GGA exchange–correlation functional[17] together with the PAW treatment of core–valence interactions[18,19]. To account the d-electron correlations, the Liechtenstein DFT+U scheme[20] was applied: a Hund's coupling $J_H$ = 1 eV[21] and on-site U parameters of 5, 6 or 8 eV for Ag. Selected calculations were also repeated using the SCAN meta-GGA[22] and the HSE06 hybrid functional[23]. In every case, the plane-wave basis was set at 520 eV. Brillouin-zone sampling used a k-point spacing of 0.032 Å⁻¹ (0.048 Å⁻¹ for HSE06) during geometry optimizations, tightening to 0.022 Å⁻¹ (0.032 Å⁻¹ for HSE06) for self-consistent-field cycles. Convergence thresholds were set to $10^{-9}$ eV for electronic iterations and $10^{-7}$ eV for ionic relaxations. Frequencies for optimized *C*2/*c* and $P\bar{1}$ structures were obtained with the PHONOPY package[24] following the detailed protocol provided on its official website. Zero-point energies and Γ-point vibrational frequencies were then extracted either via density functional perturbation theory (DFPT) or by the finite-displacement scheme implemented in VASP.
All figures of the structures were visualized with the VESTA software[25].

**RESULTS AND DISCUSSION**

*1. Structures of a novel monoclinic polymorph and of the known triclinic one from DFT*
The fully optimized crystal structures of the monoclinic and triclinic forms of $Ag_2F_5$ (using the DFT+U and HSE06 methods) are compared in **Figure 1**. In the case of the DFT+U method, calculations were performed using three values of the Hubbard parameter ($U_{Ag}$ equal to 5, 6 or 8 eV). Cell vectors and volumes are shown in **Table 1**.

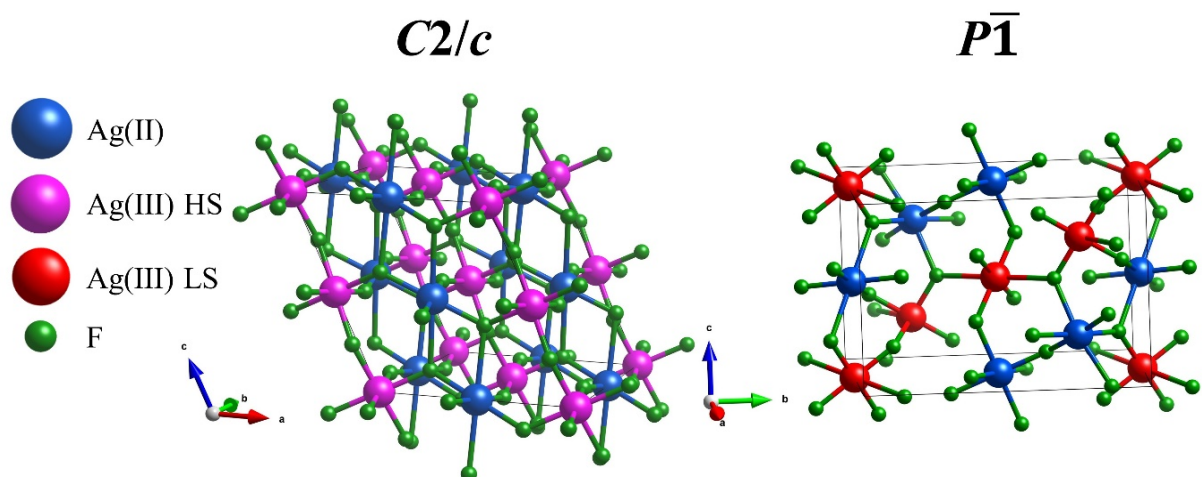

**Figure 1.** Unit cells of two polymorphic forms of Ag$_2$F$_5$: monoclinic *C*2/*c* (left) and triclinic $P\bar{1}$[12](right). LS and HS denote a low-spin or a high-spin configuration of Ag$^{3+}$, respectively. For both cells there are four formula units in the unit cell (Z = 4).

We begin the discussion with the triclinic polymorph, which has been well characterized in experiment. The geometry optimizations performed within both the SCAN and DFT+U methods, as well as the HSE06 hybrid functional yield unit cell volumes which are very close to the experimental value. In particular, the optimized cell volume is 3.49%, 2.37% and 1.08% smaller than the experimental value within the SCAN, DFT+U (U$_{Ag}$ = 5 eV) and HSE06 methods, respectively (see Table 1). This is a very good agreement especially given the fact that theoretical data correspond to $T \rightarrow 0$ K conditions, while experiments correspond to a finite temperature (hence a crystal is subject to a thermal expansion). We will focus further analysis of the Ag$_2$F$_5$ structures in both polymorphic forms on the results obtained within the reliable HSE06 method.

**Table 1.** Unit cell vectors obtained from geometry optimizations using the SCAN, DFT+U, and HSE06 methods. Experimental data are also provided for $P\bar{1}$[12] and for *C*2/*c* (following our own analysis of experimental data[11], see below). Angle values are given in Table SM.1 in SM.

| Method | | *C*2/*c* | | | | $P\bar{1}$ | | | |
|---|---|---|---|---|---|---|---|---|---|
| | | a [Å] | b [Å] | c [Å] | V [Å$^3$] | a [Å] | b [Å] | c [Å] | V [Å$^3$] |
| DFT+U | U$_{Ag}$ = 5 eV | 7.35 | 7.86 | 7.84 | 409.35 | 4.96 | 11.03 | 7.29 | 381.62 |
| | U$_{Ag}$ = 6 eV | 7.33 | 7.86 | 7.84 | 409.61 | 4.95 | 11.00 | 7.27 | 378.67 |
| | U$_{Ag}$ = 8 eV | 7.28 | 7.77 | 7.74 | 397.95 | 4.95 | 10.93 | 7.23 | 373.33 |
| SCAN | | 7.39 | 8.24 | 8.07 | 444.23 | 4.84 | 11.05 | 7.32 | 377.24 |
| HSE06 | | 7.40 | 7.99 | 7.98 | 427.62 | 4.96 | 11.07 | 7.34 | 385.66 |
| Exp.[11,12] | | 7.32-7.44 | 7.73 | 7.83-7.98 | 404.4-413.1 | 5.00 | 11.09 | 7.36 | 390.88 |

In the triclinic system ($P\bar{1}$), two types of the first coordination spheres for each cation oxidation state can be distinguished, namely two for silver(II) and two for silver(III). They differ in the relative bond lengths, bond angles, and consequently in the overall shape of the coordination sphere as described below.

The first type of Ag$^{II}$ sphere in $P\bar{1}$ (**Figure 2**, first from the left, top row) is characterized by three longer Ag–F contacts (2.25 Å, 2.36 Å, and 2.38 Å) complemented by three significantly shorter contacts (2.00 Å, 2.01 Å, and 2.02 Å), which yields a distorted octahedral geometry [AgF$_6$]$^{4-}$. The second type of first coordination sphere for silver(II) (Figure 2, top row, second from the left) adopts the shape of an elongated octahedron. Here, two axial Ag–F contacts, following HSE06 results, are equal to 2.58 Å and 2.60 Å; while the shorter equatorial Ag–F bonds have lengths of 2.11 Å, 2.10 Å, 2.05 Å, and 2.03 Å. The bond lengths and the overall shape of coordination sphere for this type of Ag$^{II}$ cations are consistent with those reported for AgF$_2$[26]. In the monoclinic system (*C2/c*), the geometry of the first coordination sphere of silver(II) can also be described as an elongated octahedron. However, in this polymorph the axial Ag–F bonds are longer than those in the triclinic system, with length 2×2.77 Å, while the equatorial Ag–F contacts are equal 2×2.00 Å and 2×2.05 Å. The distorted geometries of the [Ag$^{II}$F$_6$]$^{4-}$ coordination spheres in both polymorphs are attributed to the Jahn–Teller effect[27].

In the $P\bar{1}$ system, the first group of Ag$^{III}$ ions (Figure 2, bottom row, first from the left) exhibits a first coordination sphere with an elongated octahedral geometry, which approached a square-planar one (as typical for low-spin d$^8$ systems). The Ag–F bonds in this group have lengths of 2.73 Å and 2.76 Å, as well as 2×1.89 Å and 2×1.91 Å. The second group of Ag$^{III}$ ions forms a distorted octahedron with six F$^-$ ligands, where the F$_{ax.}$–Ag–F$_{ax.}$ angle decreases from nearly 180° to 132°, comparing to the first type of coordination spheres. In this case, the Ag–F bond lengths are equal to 2.75 Å and 2.69 Å, along with 4×(1.88–1.93 Å). In the monoclinic system, Ag$^{III}$ coordinates six fluorine atoms to form an almost ideal octahedron, the Ag–F contacts are 2×2.02 Å, 2×2.04 Å, and 2×2.05 Å, as typical for high-spin cations with a d$^8$ electronic configuration.

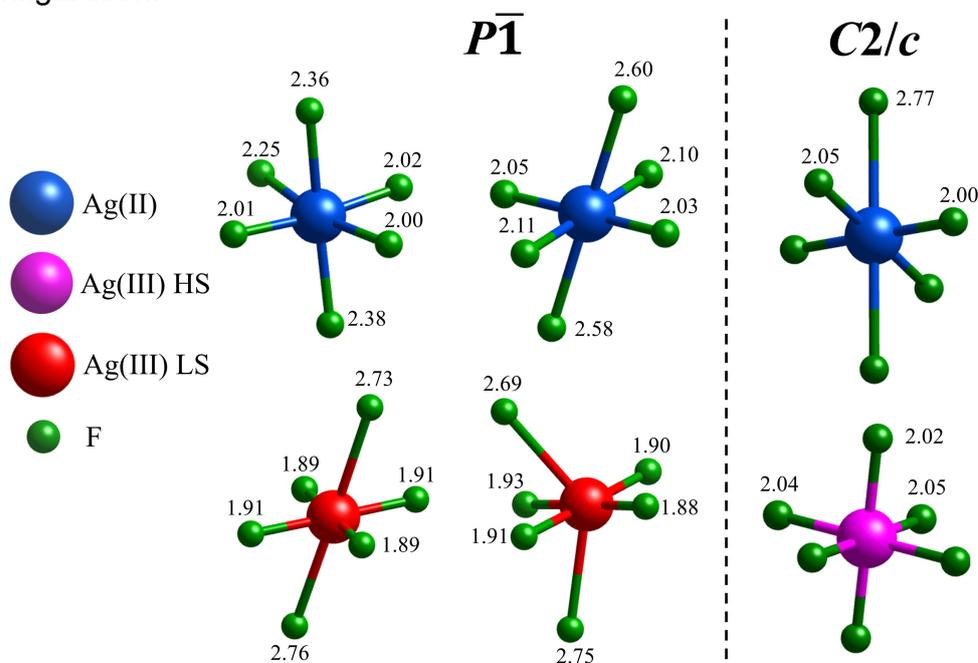

**Figure 2.** The first coordination spheres for the Ag(II) and Ag(III) ions in the triclinic and monoclinic systems. Theoretically calculated bond lengths are given in Å, based on HSE06 results.

The structural difference between coordination spheres of Ag<sup>III</sup> in both forms are consistent with the calculated magnetic moment at silver centers (0 μ$_B$ for triclinic form and 0.86μ$_B$ for monoclinic one according to the HSE06 results). Both the local geometry and magnetic moments stem from the occupation of the d($x^2$-$y^2$) and d($z^2$) orbitals: 0 + 2 electrons for low-spin and 1 + 1 electrons for high spin form, respectively.

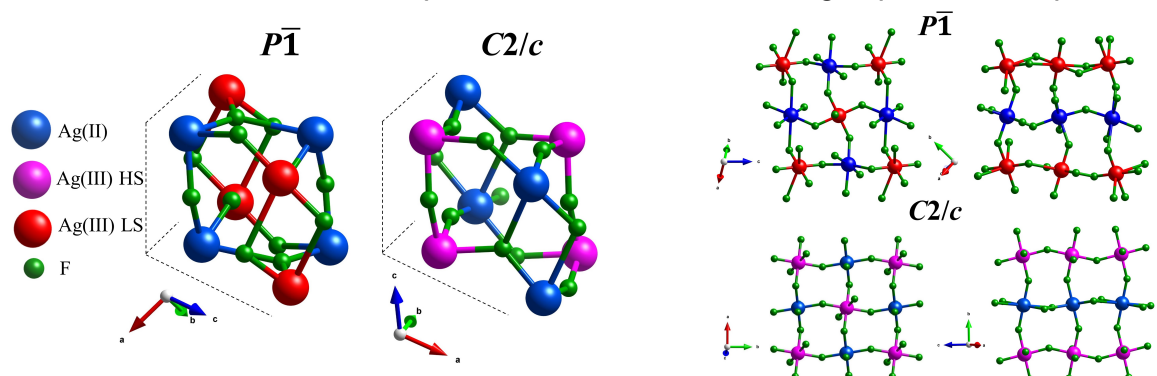

**Figure 3**. Comparison of the structural motifs and interatomic connectivity present in both structural forms of Ag$_2$F$_5$.

Despite different symmetries of both lattices, the structural motifs in both polymorphs are similar (**Figure 3**). Taking into account the interatomic connectivity it is noticeable that in both structures the Ag–F–Ag bridges connect homovalent silver cations; in the $P\bar{1}$ structure the fluoride ligands connect Ag$^{II}$ ions, whereas in the monoclinic structure they link Ag$^{III}$ ions (Figure 3, right). Therefore, at a hypothetical $P\bar{1}$ →$C2/c$ transition, the oxidation states of silver ions at corresponding crystallographic positions interchange. I.e., the site hosting Ag$^{II}$ in the triclinic structure is occupied by high-spin Ag$^{III}$ in the $C2/c$ cell (Figure 3, left), while the low-spin Ag$^{III}$ site in $P\bar{1}$ is occupied by Ag$^{2+}$ in the monoclinic form.

## 2. Lattice vibrations for two polymorphs of Ag$_2$F$_5$ from DFT

Based on group theory, one can anticipate 42 phonon vibrations for the lattice of $C2/c$ Ag$_2$F$_5$ polymorph (with Z=2 for the primitive cell). Out of these, 12 modes are silent (A$_u$), and 3 correspond to translational (acoustic) modes (2B$_u$ and A$_u$). 12 modes are infrared active (12B$_u$), while 15 modes are Raman active (8B$_g$ + 7A$_g$) (see **Table 2**).

**Table 2** Phonon vibration frequencies and their symmetries from DFT+U calculations for $C2/c$ polymorph. Activity "IR" stems for infrared spectra, "R" for Raman ones and "silent" correspond to inactive bands.

| # | Symmetry | Frequency [cm$^{-1}$] | Activity | # | Symmetry | Frequency [cm$^{-1}$] | Activity |
|---|---|---|---|---|---|---|---|
| 1 | B$_u$ | 542 | IR | 22 | B$_u$ | 185 | IR |
| 2 | A$_u$ | 537 | silent | 23 | A$_u$ | 169 | silent |
| 3 | B$_u$ | 536 | IR | 24 | A$_u$ | 162 | silent |
| 4 | A$_u$ | 467 | silent | 25 | A$_g$ | 160 | R |
| 5 | B$_u$ | 450 | IR | 26 | B$_u$ | 159 | IR |
| 6 | B$_g$ | 449 | R | 27 | A$_u$ | 153 | silent |

| 7  | $A_g$ | 432 | R      | 28 | $B_u$ | 148 | IR     |
|----|-------|-----|--------|----|-------|-----|--------|
| 8  | $B_g$ | 428 | R      | 29 | $A_u$ | 140 | silent |
| 9  | $A_g$ | 393 | R      | 30 | $B_u$ | 125 | IR     |
| 10 | $B_g$ | 377 | R      | 31 | $B_g$ | 124 | R      |
| 11 | $A_u$ | 343 | silent | 32 | $A_g$ | 97  | R      |
| 12 | $B_u$ | 324 | IR     | 33 | $B_u$ | 86  | IR     |
| 13 | $A_u$ | 282 | silent | 34 | $B_g$ | 78  | R      |
| 14 | $A_u$ | 272 | silent | 35 | $A_u$ | 76  | silent |
| 15 | $B_u$ | 267 | IR     | 36 | $A_g$ | 65  | R      |
| 16 | $B_g$ | 252 | R      | 37 | $A_u$ | 55  | silent |
| 17 | $A_g$ | 242 | R      | 38 | $B_g$ | 46  | R      |
| 18 | $A_g$ | 206 | R      | 39 | $B_u$ | 46  | IR     |
| 19 | $B_g$ | 198 | R      | 40 | $B_u$ | 0   | IR     |
| 20 | $A_u$ | 197 | silent | 41 | $B_u$ | -1  | IR     |
| 21 | $B_u$ | 194 | IR     | 42 | $A_u$ | -1  | silent |

The acoustic modes have null frequencies with a small discrepancy up to 1 cm$^{-1}$. There are no imaginary modes which testifies that the monoclinic polymorph is a genuine minimum on the potential energy surface, and as such is synthesizable. The optical modes stretch up to 542 cm$^{-1}$ which is a typical value for the Ag–F stretching modes in Ag(II) compounds. A similar analysis for the triclinic polymorph (**SM V**), again suggests the absence of any imaginary phonons. The triclinic form has as many as 81 optical modes altogether (with no silent modes), plus three acoustic phonons. The highest (Debye) frequency computed is 584 cm$^{-1}$. The IR and Raman spectra for both forms will differ considerably, so they could be easily detected in experiments based on the characteristic (fingerprint) vibrations. Regretfully, the vibrational spectra have not been measured for any of the previously reported samples[11,12] but our theoretical values could prove to be useful if these experiments are repeated in the future.

Having described structure and lattice dynamics of the known triclinic and the novel monoclinic form of Ag$_2$F$_5$ proposed here, we now confront these results with the experimental data available in the literature.[11,12]

*3. The analysis of the powder x-ray diffraction data available in the literature*
At this juncture, it is worthwhile to introduce findings from the two experimental studies that employed entirely different synthesis conditions to obtain Ag$_2$F$_5$. Given the marked differences in the synthetic approaches between the works of Žemva[11] and Fischer[12], a detailed analysis of the X-ray diffractions presented in both works is of particular interest. Below we show comparison of their diffractograms; the one from Fischer[12], is a theoretically generated pattern for their triclinic structure (assuming Cu radiation) while that from Žemva[11] contains only the reported positions and intensities of reflections (blue dots). We have also included the one generated for theoretical structure of monoclinic Ag$_2$F$_5$ (*C2/c*) system based on DFT+U calculations (black curve) (**Figure 4**).

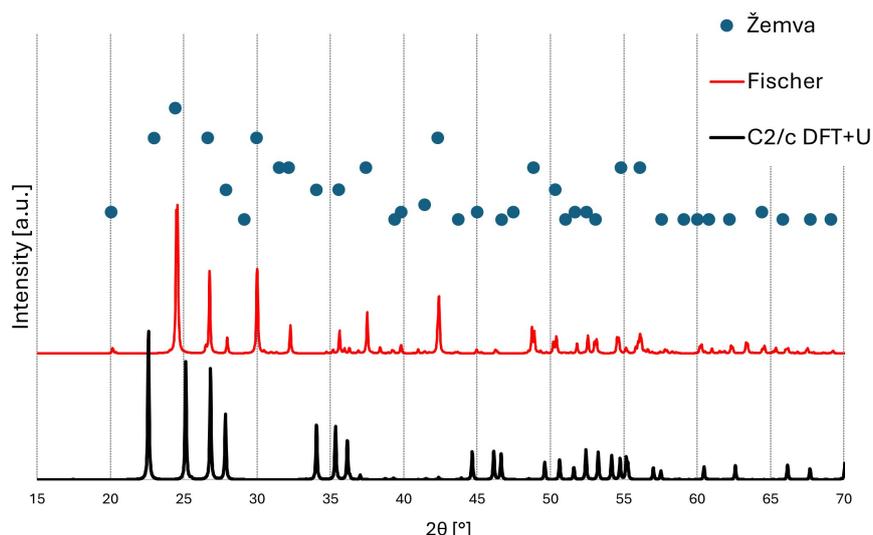

**Figure 4.** Comparison of $Ag_2F_5$ X-ray patterns: (i) as reported by Žemva[11] (blue dots), (ii) generated from the experimental triclinic structure[12] (red line), (iii) generated from the monoclinic structure (DFT+U calculations, $U_{Ag}$ = 5eV) (black line).

The comparison reveals a significant difference in the number and positions of reflections measured for the polycrystalline sample obtained by Žemva[11] and single crystal measured by Fischer[12]. Although the triclinic $Ag_2F_5$ is certainly present in Žemva's sample, yet these authors have observed many more unique reflections which cannot be explained by the presence of $KAsF_6$ byproduct nor of any other known Ag–F system. Analysis of these additional reflections (**Table 3**) reveals a similarity of their positions to the ones expected for the monoclinic $C2/c$ form of $Ag_2F_5$.

**Table 3.** Positions (2θ) of reflexes, spacing between crystallographic planes and plane indexes for $Ag_2F_5$ in both, $C2/c$ (theoretical, DFT+U method) and $P\bar{1}$, forms. Positions of the reflections observed by Žemva et al.[1] are also presented. "NO" means "not observed". Angles correspond to Cu lamp radiation.

| Žemva[11] | | $Ag_2F_5$ ($C2/c$)$^{theoretical}$ | | | $Ag_2F_5$ ($P\bar{1}$)[12] | | |
|---|---|---|---|---|---|---|---|
| 2θ [°] | d [Å] | 2θ [°] | d [Å] | index | 2θ [°] | d [Å] | index |
| 20.08 | 4.42 | NO | NO | NO | 20.10 | 4.41 | (1-10) |
| 23.00 | 3.86 | 22.60 | 3.93 | (020) | NO | NO | NO |
| 24.43 | 3.64 | 25.12 | 3.54 | (002) | 24.58 | 3.62 | (120) |
| 26.66 | 3.34 | 26.82 | 3.32 | (200) | 26.76 | 3.33 | (10-2) |
| 27.89 | 3.20 | 27.84 | 3.20 | (20-2) | 27.96 | 3.19 | (1-1-2) |
| 29.16 | 3.06 | NO | NO | NO | NO | NO | NO |
| 29.99 | 2.98 | NO | NO | NO | 30.00 | 2.98 | (022) |
| 32.19 | 2.78 | NO | NO | NO | 32.22 | 2.78 | (040) |
| 34.05 | 2.63 | 34.04 | 2.63 | (022) | NO | NO | NO |
| 35.60 | 2.52 | 35.35 | 2.54 | (220) | 35.62 | 2.52 | (102) |
| 37.45 | 2.40 | 37.02 | 2.43 | (22-2) | 37.52 | 2.40 | (200) |
| 41.43 | 2.18 | NO | NO | NO | 41.44 | 2.18 | (042) |
| 45.02 | 2.01 | 44.66 | 2.03 | (202) | 44.98 | 2.01 | (2-30) |
| 46.68 | 1.94 | 46.64 | 1.95 | (20-4) | NO | NO | NO |

For example, the strong reflection observed at *ca.* 23° is very similar to the most intense (020) reflex characteristic of the theoretically predicted monoclinic Ag$_2$F$_5$ structure – around 22.6°. Similarly, the reflection seen at *ca.* 27.9° cannot be explained solely by the (quite weak) (1-1-2) reflex coming from the $\overline{P1}$, form but it may also stem from the strong (20-2) reflex of the *C*2/c cell, etc. Looking at the low-angle part of the diffractogram (where the measured intensities are the strongest) we were able to identify several strong reflections in the pattern published by Žemva[11] which may be explained by the presence of the monoclinic polymorph in their samples. Fortunately, the (hkl) indexes for these reflections are not redundant and we were able, based on their experimental positions, to estimate approximate unit cell vectors for the *C*2/c cell coming from our analysis of the previous experiments. We have used several independent sets of the four reflections needed to calculate three lattice constant parameters and a β angle, and the obtained sets of lattice parameters have been listed in **Table 4**. The corresponding generated powder patterns are shown on **Fig SM.1 in SM**.

**Table 4**. The four sets of selected reflections observed by Žemva et al.**[1]** used for calculations of lattice parameters of the monoclinic form of for Ag$_2$F$_5$ together with the resulting lattice parameters, β angle, and unit cell volume, V. Ranges of theoretical values have also been provided.

| | Selected reflexions from Žemva | | Estimated lattice parameters and unit cell volume | | | | |
|---|---|---|---|---|---|---|---|
| Set | d [Å] | hkl | a [Å] | b [Å] | c [Å] | β [°] | V [Å$^3$] |
| I | 3.86, 3.64, 3.34, 2.01 | 020, 002, 200, 202 | 7.71 | 7.73 | 8.40 | 119.9 | 433.4 |
| II | 3.86, 3.34, 3.20, 1.94 | 020, 200, 20-2, 20-4 | 7.37 | 7.73 | 7.83 | 114.9 | 404.4 |
| III | 3.86, 3.20, 2.63, 1.94 | 020, 20-2, 02-2, 20-4 | 7.44 | 7.73 | 7.85 | 113.7 | 413.1 |
| IV | 3.86, 3.64, 3.34, 3.20 | 020, 002, 200, 20-2 | 7.32 | 7.73 | 7.98 | 114.1 | 411.7 |
| Ranges of theoretical values (except for SCAN results) | | | 7.28–7.40 | 7.77–7.99 | 7.74–7.98 | 114.8-115.3 | 398.0–427.6 |

It seems that set I of four reflections is incorrect, as it may be judged from lattice parameters a and c too much departing from the values expected from theoretical calculations. However, sets II, III and IV all give lattice parameters and volumes which are reasonably similar to the values predicted by theory. Therefore, the associated ranges of the lattice parameter values from sets II-IV have been introduced to the previously discussed Table 1. Regretfully, since Žemva et al. did not provide a full powder pattern, we could not perform any fits to see whether any of the sets II-IV performs better than the others in the Rietveld fit. Nevertheless, the abovementioned results suggest that the sample obtained by Žemva was, in fact, a polycrystalline mixture. It consisted of a triclinic Ag$_2$F$_5$ (only eleven years later characterized by Fischer), and the then-overlooked monoclinic Ag$_2$F$_5$. Importantly, the diffractogram reported by Žemva contains also additional reflexes that are absent in the diffraction patterns of both considered polymorphic forms of Ag$_2$F$_5$ (e.g., at *2θ* of 37.2° and 42.3°). This may imply that at least three phases were present in Žemva's sample, which naturally explains their difficulties to solve the crystal structure for such complex sample. The fact that these authors[11] did not report magnetic susceptibility data for their sample of $Ag_2F_5$ is also characteristic, as they characterized all other compounds in their study using this method.

Taking into account the coexistence of two polymorphs of $Ag_2F_5$ in Žemva's sample, the following section will address the relative stability of these structural forms.

*4. Relative thermodynamic and energetic stability of both polymorphs of $Ag_2F_5$*

The energy of $Ag_2F_5$ formation (*dE,* equation 2) was calculated by considering a synthesis reaction between the binary systems, $AgF_2$ and $AgF_3$ (equation 1). Additionally, the change in the volume of the reactants ($V_{AgF_2} + V_{AgF_3}$) relative to the product ($V_{Ag_2F_5}$) was calculated (*dV*, equation 3). All calculations were performed for $U_{Ag}$ values of 5, 6, and 8 eV within the DFT+U framework. The obtained *dE* and *dV* values for both structural forms of $Ag_2F_5$ and for the three Hubbard $U_{Ag}$ parameters are summarized in the table below (**Table 5**).

$$AgF_2 + AgF_3 \rightarrow Ag_2F_5 \quad \text{(eq. 1)}$$
$$dE = E_{Ag_2F_5} - (E_{AgF_2} + E_{AgF_3}) \quad \text{(eq. 2)}$$
$$dV = V_{Ag_2F_5} - (V_{AgF_2} + V_{AgF_3}) \quad \text{(eq. 3)}$$

Here, the values of *E* for the individual systems represent the <u>electronic plus spin</u> total energy in their magnetic ground state, as obtained from DFT+U calculations. It should be noted that $AgF_2$ substrate in its ground state exhibits strong antiferromagnetic coupling between silver(II) ions within the layers[1,30]. In contrast, $AgF_3$ in its ground state contains low-spin $Ag^{III}$, therefore it is diamagnetic.

**Table 5**. Energy effect (*dE*, equation 2) and differential volume (*dV*, equation 3) of the formation reaction of $Ag_2F_5$ in two polymorphic forms. Values are given per $Ag_2F_5$ formula unit (FU). The *-dS* parameter was determined using a semi-empirical method, where $dS/dV$ = 18.24 meV K$^{-1}$ FU$^{-1}$ per 1 nm³ of volume change in the system[31]. The temperature *T* in term *-TdS* was assumed to be 298 K.

| DFT + U $U_{Ag}$ parameter | dE [meV/FU] | | dV [Å³] | | -TdS [meV/FU] | | dE-TdS [meV/FU] | |
|---|---|---|---|---|---|---|---|---|
| | $P\bar{1}$ | C2/c | $P\bar{1}$ | C2/c | $P\bar{1}$ | C2/c | $P\bar{1}$ | C2/c |
| 5 eV | -69 | -85 | -0,69 | +6,24 | +4 | -34 | -65 | -119 |
| 6 eV | -66 | -86 | -0,97 | +6,76 | +5 | -37 | -61 | -123 |
| 8 eV | -62 | -99 | -0,89 | +8,18 | +5 | -44 | -57 | -143 |

The energy effect (*dE*) is negative and quite substantial for both types of $Ag_2F_5$ (from –62 to –99 meV per FU), regardless of the value of $U_{Ag}$ parameter. This alone suggests that both forms might be prepared in experiment. However, in the case of the *C2/c* structure, the value of the formation energy is more negative than for the triclinic form by ca. 16–37 meV per FU. Thus, the energy term stabilizes the monoclinic structure, the triclinic one being a metastable form at T → 0 K.

Significant differences in *dV* values are also computed between the two structural forms of $Ag_2F_5$. In the triclinic system, the unit cell volume is only slightly <u>smaller</u> than the sum of the unit cell volumes of $AgF_2$ and $AgF_3$, ranging from 0.69 to 0.89 Å³ per formula unit (FU), depending on the chosen $U_{Ag}$ parameter. In contrast, the unit cell volume of the monoclinic system is 6.24 to 8.18 Å³ larger than the sum of the unit cell volumes of precursors (eq. 1). Given the substantial differences in volumetric effects between the synthesis reactions of monoclinic and triclinic $Ag_2F_5$, the entropic contribution was estimated using a semi-empirical method, where volume expansion (*dV*) is linked to entropy increase (*dS*)[21]. The negative *dS* value for the synthesis of the triclinic phase results in a positive *TdS* contribution to the Gibbs free energy expression. At room temperature is approximately +4 to +5 meV/FU. This effect becomes increasingly significant as the temperature rises. Consequently, considering only this contribution to the entropy, the formation of the triclinic phase from the binary substrates becomes progressively less favorable. For this reason, it is not surprising that under Fischer's synthesis conditions of T = 720K[12], reproducibility issues were encountered both in the synthesis reaction and in the isolation of crystals of the triclinic form.

Conversely, the positive *dS* of formation for the monoclinic $Ag_2F_5$ phase indicates the opposite tendency. Assuming a temperature of 298 K (room temperature), the entropy contribution (-*TdS*) is expected to considerably stabilize the monoclinic form, reducing the Gibbs free energy by approximately -34 or -44 meV/FU, depending on the chosen $U_{Ag}$.

It should be noted, that temperature is not the only factor influencing the relative stability of the two structural forms of $Ag_2F_5$. The conditions of elevated pressure (*p* = 0.04 GPa[12] was used by Fischer) must also be considered. Our results show, that with increasing pressure, the triclinic form becomes energetically more favorable (see *SM III*). This is evident in the energy difference ($E_{P\bar{1}}$ - $E_{C2/c}$) between the structural forms as pressure increases (exact values are presented in SM, Fig SM.2). At p = 2 GPa, $E_{P\bar{1}}$ - $E_{C2/c}$ is –83 meV/FU, and at 10 GPa, it reaches -446 meV/FU, indicating that the triclinic form is increasingly favored under higher pressures. Preference for the low-volume triclinic form under elevated pressure is of course expected taking into account the pV term.

The analysis suggests that the triclinic ($P\bar{1}$) $Ag_2F_5$ structure reported in the literature is energetically less stable than the *C2/c* one under conditions of *p* → *0* and *T* → *0*. Monoclinic form enhances its stability as T increases, while the triclinic one could become a ground state at elevated pressure.

The above-mentioned calculations do not account for the contribution of lattice vibrations, as these are usually negligible for elements heavier than boron. Nevertheless, we performed zero-point energy (ZPE) calculations to evaluate phonons' contribution to the total energy ($E_{ZPE}$) and the relative stability of the structures ($dE_{ZPE}$), considering *p* → *0* conditions. The comparison of calculation results is presented in the **Table 6** below.

Table 6 presents selected reaction parameters for the hypothetical structural transformation reaction of $Ag_2F_5$ from the monoclinic to the triclinic form (eq. 4). The energy difference between the two $Ag_2F_5$ forms was determined both without (*dE*) and with ($dE_{ZPE}$) the inclusion of zero-point vibrational energy (eq. 5), along with the volumetric effect of the reaction (eq. 6).

$$Ag_2F_{5(C2/c)} \rightarrow Ag_2F_{5(P\bar{1})} \quad \text{(eq. 4)}$$
$$dE = E_{P\bar{1}} - E_{C2/c} \quad \text{(eq. 5)}$$
$$dV = V_{P\bar{1}} - V_{C2/c} \quad \text{(eq. 6)}$$

The entropy change arising from the volume difference between to $Ag_2F_5$ systems (eq.6) was estimated using a semi-empirical method, where $dS/dV$ = 18.24 meV·K$^{-1}$·FU$^{-1}$ per 1 nm³ of volume change[21]. By including the entropic contribution associated with volume expansion, the equilibrium temperature for the coexistence of both phases was determined according to the equation 7 below.

$$T_{eq} = \frac{dE_{ZPE}}{dS} \quad \text{(eq. 7)}$$

One may also compute the corresponding equilibrium pressure:

$$p_{eq} = \frac{dE_{ZPE}}{dV} \quad \text{(eq. 8)}$$

**Table 6** Parameters $dE$, $dE_{ZPE}$ (eq. 5), the entropic effect ($dS$), and the equilibrium temperature (eq. 7) and pressure (eq. 8).

| Parameter | DFT + U | |
| --- | --- | --- |
| | U = 5 | U = 8 |
| dE [meV/FU] | +16 | +37 |
| dE$_{ZPE}$ [meV/FU] | +36 | +38 |
| dV [Å³/FU] | -6.93 | -6.15 |
| dS [meV/K·FU] | -0.126 | -0.112 |
| T$_{eq}$ [K] | -285 | -339 |
| p$_{eq}$ [GPa] | 0.83 | 0.99 |

It turns out that phonons' contribution to dE is ca. 1 meV (U = 8eV) or 20 meV (U = 5 eV); however, relative stability of polymorphs is not qualitatively altered as compared to results which do not take phonons into account. This is reflected in the formally <u>negative</u> value of $T_{eq}$ (equation 7), which considers both entropy change ($dS$) and the energy difference including contribution from zero-point vibrations ($E_{ZPE}$). In other words, this indicates that the $P\bar{1}$ form should not be favored at any finite temperature. On the other hand, the triclinic form could be stabilized at slightly elevated pressure of the order of 0.8–1.0 GPa for T → 0 conditions.

Interestingly, considering DFT+U calculations without spin polarization, the energy difference between the triclinic and monoclinic structures is -141 meV/FU in favor of the former. This suggests that magnetic interactions between spins play a dominant role in favoring $C2/c$ over the $P\bar{1}$. This is a highly unusual situation, as most often the contribution of magnetic interactions to the free energy of polymorphic transformation reaction is negligible; here, the low-to-high spin transformation for Ag(III) allows for the appearance of many new superexchange (SE) pathways characterized by strong SE coupling constants, as it will be detailed in the next section.

*5. Electronic and magnetic properties*

**Figure 5** illustrates the electronic density of states for the two structural forms of $Ag_2F_5$. The value of band gap between the occupied and unoccupied states is very similar in both polymorphs. For a Hubbard parameter of $U_{Ag}$ = 5 eV, the *C2/c* phase exhibits a gap of 0.73 eV, while the $P\bar{1}$ phase has a band gap of 0.74 eV. When $U_{Ag}$ is increased to 8 eV, the band gap increases to 1.47 eV for the *C2/c* phase and to 1.27 eV for the $P\bar{1}$ phase. For both polymorphs, a significant similarity is observed in the character of the valence band near the Fermi level, where the dominant contribution arises from fluorine *p* states. In contrast, the conduction band is primarily composed of unoccupied metal *d* states. Consequently, the band gap in both structures is of the charge-transfer (CT) type, according to Zaanen-Sawatzky-Allen (ZSA) classification[32].

However, in the *C2/c* system, the conduction band exhibits a pronounced separation between the silver(II) and silver(III) states. The unoccupied states closest to the Fermi level (at 0 eV; see **Figure 5**) arises from the $d_{(x^2-y^2)}$ and $d_{(z^2)}$ high-spin $Ag^{III}$ states; while the Ag $d_{(x^2-y^2)}$ states of $Ag^{II}$ are located by 1 eV higher in energy scale. This observation suggests that HS $Ag^{III}$ exhibits a more oxidizing character than $Ag^{II}$. This conclusion is significant for potential electron-doping of the monoclinic polymorph, as it suggests that additional electron density would preferentially populate the $Ag^{III}$ $d_{(x^2-y^2)}$ and/or $d_{(z^2)}$ rather than the $Ag^{II}$ $d_{(x^2-y^2)}$ states. Essentially, the *C2/c* structure could facilitate the attainment of an intermediate oxidation state between II and III for silver cation – a requirement postulated as key to achieving superconductivity in Ag-F systems[7,9].

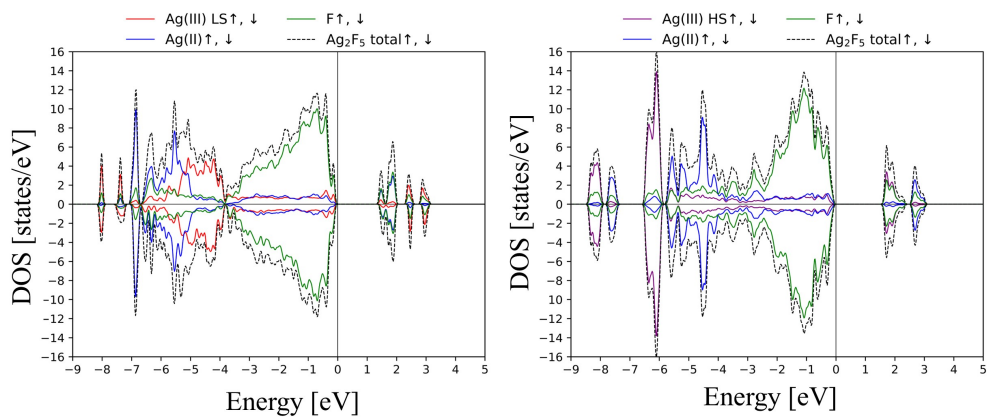

**Figure 5** Comparison of the density of states (DOS) for the $P\bar{1}$ (left) and *C2/c* (right) structures. "LS" denotes the low-spin state, while "HS" denotes the high-spin state. Results within DFT+U, where $U_{Ag}$ = 8 eV[3].

For the triclinic $P\bar{1}$ phase, the conduction band is predominantly composed of unoccupied Ag $d_{(yz)}$ states from both silver(II) and silver(III). The calculated magnetic moment LS-$Ag^{III}$ cation in $P\bar{1}$ form equals 0 $\mu_B$, as expected, so entire magnetism of this phase comes from $Ag^{II}$. Unfortunately, despite multiple attempts, Fisher and Muller were unable to obtain sufficiently pure $Ag_2F_5$ samples required for magnetic analysis[12]. Based on the calculation we propose, that this compound can be regarded as a diluted magnet with rather weak magnetic interactions between $Ag^{II}$ cations due to the low angle of the $Ag^{II}$-F-$Ag^{II}$ bridge (128°) and the alternating Ag-F distances within the linkage (2.1 Å/1.9 Å). For this reason, in the following part only magnetic properties of *C2/c* form will be described.

DFT+U calculations conducted for the *C2/c* phase using a cell presented in **Figure 6** reveal that the magnetic ground state corresponds to a state designated as AFM2 (see *Supplementary Materials*); same as the one proposed for isostructural $AgCoF_5$ compound[15]. Therefore, the magnetic structure of $Ag_2F_5$ can be classified as a G-type antiferromagnet (see *Supplementary Materials*).

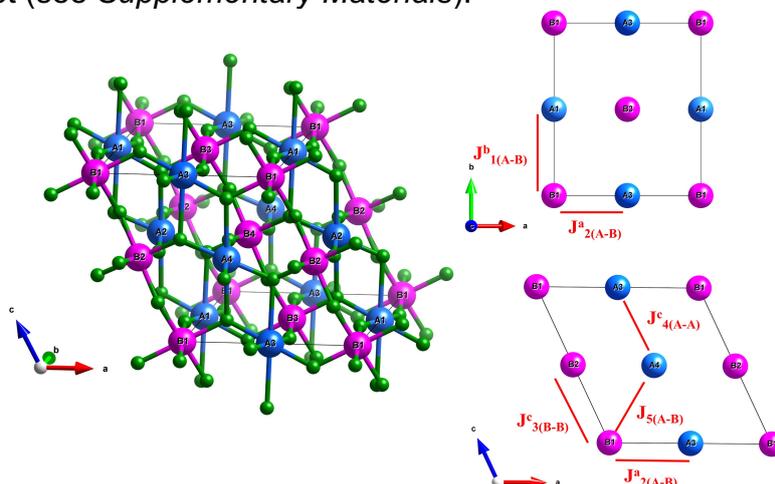

**Figure 6**. The $Ag_2F_5$ cell adapted for superexchange constant calculations, with projections onto the (001) and (010) crystallographic planes and superexchange pathways ($Ag^{II}$ – blue, $Ag^{III}$ – pink, F – green).

To examined the magnetic interaction strengths in new polymorph of $Ag_2F_5$ we calculated five superexchange constants (*J*) using the Heisenberg Hamiltonian, based on the energies of the corresponding spin configurations (broken-symmetry method) within DFT+U method. The Hamiltonian employed for the calculation of the superexchange constants is presented in *Supplementary Materials*.

The calculated superexchange constants, together with the most relevant structural parameters – the distance between the centers of the paramagnetic species and the F-Ag-F bond angle specific to each exchange pathway – are presented in **Table 7** below.

It is worth to note, that the HSE06 method for reference system, $AgF_2$, yields the superexchange constant reaches 75% of the experimental value[1], indicating good agreement with experiment. In the *C2/c* structure of $Ag_2F_5$ a pronounced anisotropy in the strength of magnetic interactions, dependent on the crystal direction, is observed. The superexchange constant between $Ag^{III}$ and $Ag^{II}$ is –186 meV (along *b* direction, $J^b_{1(A-B)}$, where A=$Ag^{II}$, B=$Ag^{III}$), while the second strongest coupling, $J^c_{4(A-A)}$, which characterizes the magnetic interaction between $Ag^{III}$ ions, is –93 meV (along *c* direction, $U_{Ag}$ = 5 eV). In both cases, the A-F-B and B-F-B bridge angles along specific pathways (1 and 4) are among the closest to 180° (154°/153°) of all five pathways, which favors antiferromagnetic interactions according to the Goodenough-Kanamori-Anderson rules[33–35].

**Table 7** Superexchange coupling constant values determined using DFT+U methods, along with the corresponding crystallographic directions, bond angles, and distances between the paramagnetic cations relevant to each exchange pathway. Negative J values correspond to antiferromagnetic interaction.

| Direction | Parameter (A = $Ag^{2+}$, B = $Ag^{3+}$) | DFT+U ($U_{Ag}$= 5 eV) | DFT+U ($U_{Ag}$= 6 eV) | DFT+U ($U_{Ag}$= 8 eV) |
|---|---|---|---|---|
| [010] | $J^b_{1(A-B)}$ [meV] | **-186** | **-137** | **-101** |
|  | d [Å] | 3.93 | 3.93 | 3.88 |
|  | angle [°] | 154.0 | 153.2 | 152.3 |
| [100] | $J^a_{2(A-B)}$ [meV] | **-51** | **-24** | **-22** |
|  | d [Å] | 3.67 | 3.66 | 3.64 |
|  | angle [°] | 126.5 | 126.4 | 126.0 |
| [001] | $J^c_{3(B-B)}$ [meV] | **-93** | * | **-72** |
|  | d [Å] | 3.91 | 3.92 | 3.87 |
|  | angle [°] | 153.0 | 153.7 | 151.1 |
| [001] | $J^c_{4(A-A)}$ [meV] | **-13** | * | **±0** |
|  | d [Å] | 3.91 | 3.92 | 3.87 |
|  | angle [°] | 110.7 | 110.3 | 109.6 |
| [101] | $J_{5(A-B)}$ [meV] | **-7** | **-5** | **-2** |
|  | d [Å] | 4.06 | 4.09 | 4.05 |
|  | angle [°] | 117.7 | 118.4 | 119.1 |

* the exact value was not determined because the required magnetic states did not converge satisfactorily well

To provide a broader context for our findings, **Table 8** sets the three most significant (in terms of their absolute value) superexchange constants calculated for representative homodimetallic(II/III) pentafluoride systems. Two of them, $Cu_2F_5$[36,37] and $Ni_2F_5$[38], have been proposed as potentially novel compounds featuring high-spin TM-cations with mixed valence (II/III); however, neither has yet been experimentally obtained. Additionally, Table 8 presents the superexchange constants in the hypothetical compound $Ag^{II}Cu^{III}F_5$, which, according to the DFT+U calculations[15], is energetically stable relative to potential substrates, $AgF_2$ and $CuF_3$[36]. We also calculated J values for $Ag_2F_5$ using a CPU-demanding HSE06 method. This method suggests that one-dimensional antiferromagnetic interaction between mixed-valence silver centers, $J^b_{1(A-B)}$, to reach impressive –207 meV (see Table 8).

The J values indicating the strength of A-B and B-B interactions in $Cu_2F_5$, $Ni_2F_5$ and $AgCuF_5$ are several times smaller in absolute terms, compared to analogous values for $Ag_2F_5$. This suggests that the mixed-valent Ag-F system is the most promising candidate among those HTSC precursors, while taking into account J vs. $T_c$ correlation in parent undoped cuprates[39]. The monoclinic form of $Ag_2F_5$ discussed here belongs to a small group magnetic materials with exceptionally high superexchange constants which may surpass the $J_{2D}$ and $J_{1D}$ values observed for a number of experimentally-characterized cuprates[40].

**Table 8** Comparison of the three strongest (by absolute value) magnetic superexchange constants for the hypothetical compounds $Ag_2F_5$, $Cu_2F_5$, $Ni_2F_5$, and $AgCuF_5$. A – divalent metal cation, B – trivalent metal cation.

| $A^{II}B^{III}F_5$ | $U_M$ | $J^b_{1(A-B)}$ [meV] | $J^a_{2(A-B)}$ [meV] | $J^c_{3(B-B)}$ [meV] |
|---|---|---|---|---|
| $Ag^{II}Ag^{III}F_5$ | 5 eV | -186 | -51 | -93 |
|  | 8 eV | -101 | -22 | -72 |
|  | HSE06 | -207 | -50 | -87 |
| $Cu^{II}Cu^{III}F_5$ | *4 eV[37] | -33 | -7 | -34 |
|  | *6 eV[37] | -35 | -7 | -40 |
|  | **10 eV | -39 | -10 | -33 |
| $Ni^{II}Ni^{III}F_5$ | 6 eV | -12 | -5 | -11 |
| $Ag^{II}Cu^{III}F_5$ | Ag, Cu = 5 eV | -111 | -27 | -48 |

* *C*2/*m* form following Rybin et al.[36], ** *C*2/*c* form from this work with HS cation B[1].

**Conclusions**

Our results suggest the conceivable existence of $Ag_2F_5$ in the monoclinic *C*2/*c* polytype. This form features rare[41] HS-$Ag^{III}$ cations in the octahedral ligand environment rather than the more common LS ones (in the square-planar ligand field). Our calculations predict that the new form is slightly more stable than the known triclinic one at p → 0 conditions and at any temperature; the triclinic form featuring LS $Ag^{III}$ has a smaller molar volume than the monoclinic one and as such is preferred under elevated pressure of ca. 1 GPa. Consequently, the triclinic form accidentally prepared by Fischer (during crystallization attempts of $AgRhF_6$ at *p* = 400 bar, *T* = 720 K)[13], and further characterized at ambient (p,T) conditions, may have formed as a metastable phase. On the other hand, Žemva and Muller[11], employing synthesis conditions distinct from those of Fischer, did not unequivocally identify all the constituents of the final powder. Our analysis of the reflex positions in the X-ray diffractogram reported by these authors indicates the presence of two distinct polymorphs ($P\bar{1}$ and *C*2/c) in their samples, together with an unknown impurity. As these authors used the source of LS $Ag^{III}$ for their synthesis ($KAgF_4$), the formation of some amount of the *C*2/*c* form suggests that it is more stable at synthesis conditions that the triclinic one, as LS-to-HS transition has occurred to some extent (supposedly hindered only by insufficiently fast kinetics). This phenomenon is described in the literature as spin-crossover[18,19], and broadly represented by many $d^8$ $Ni^{2+}$ systems, among others. However, the LS-to-HS transition for isoelectronic $Ag^{III}$ has not yet been explicitly mentioned in the literature, and the case described here seems to be the first of its kind.

---

[1] Interestingly our calculations reveal that the *C*2/c polymorph $Cu_2F_5$ is by -56 meV/FU more stable, following DFT+U with U = 10 eV, than previously proposed *C*2/m one.

Interestingly, following our quantum mechanical calculations, the monoclinic form turns out to be less stable than the triclinic one if magnetic interactions are not accounted for. Therefore, magnetic interactions are a pivotal determinant in stabilizing the monoclinic form. This is because monoclinic form hosts very strong magnetic superexchange interactions between $d^9$ and $d^8$ silver cations mediated by fluorine ligand (which are absent for the triclinic form with LS Ag$^{III}$). The SE constants for this interaction may exceed –200 meV as calculated using HSE06 functional. The monoclinic $Ag_2F_5$ is therefore a very rare case when magnetic interactions decide the fate of relative stability of diverse polymorphic forms. Moreover, taking into account the immensely strong magnetic interaction between $d^9$ and $d^8$ silver cations mediated by fluorine ligand, and simultaneously the strong $d^9 \ldots d^9$ interactions, a potential doping of the *C*2/*c* phase may result in superconductivity with a high critical temperature.

**Acknowledgements**
The authors acknowledge the Polish National Science Center (NCN) for the project M(Ag)NET (2024/53/B/ST5/00631), and the Interdisciplinary Centre for Mathematical and Computational Modelling, University of Warsaw (ICM UW) for the project SAPPHIRE (GA83-34).

SUPPLEMENTARY MATERIALS
to
# Magnetic interactions as a pivotal determinant in stabilizing a novel $Ag^{II}Ag^{III}F_5$ polymorph with high-spin $Ag^{III}$

Daniel Jezierski[1*] and Wojciech Grochala[1*]

[1] *Center of New Technologies, University of Warsaw, 02089 Warsaw, Poland*
*\*d.jezierski@cent.uw.edu.pl, w.grochala@cent.uw.edu.pl*

## I.   Diffractogram analysis

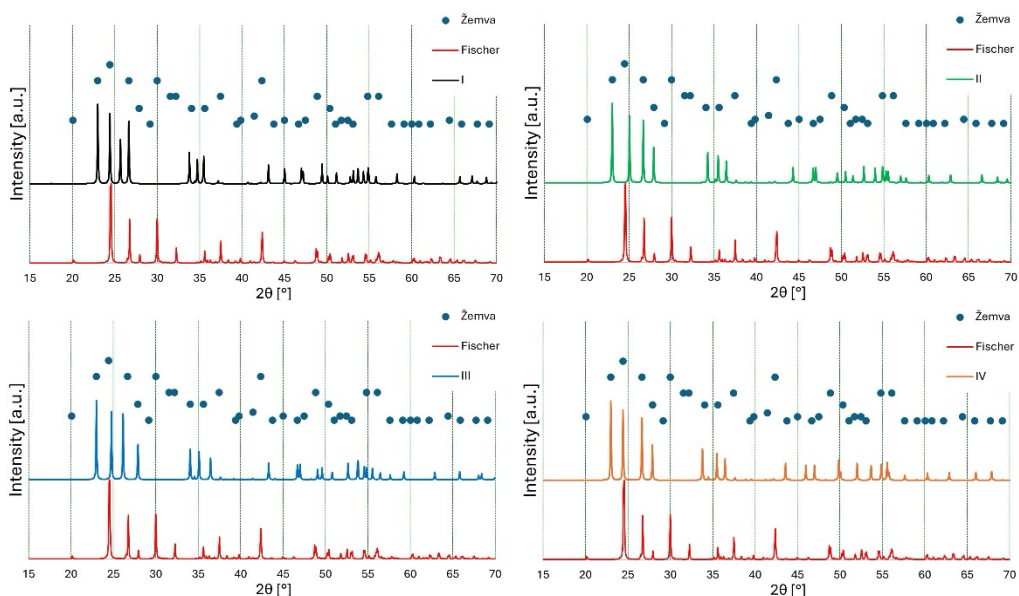

**Fig SM.1** XRD patterns comparison – I, II, III and IV for independent sets of reflexes.

## II.   Supplementary structural data for the two polymorphs

**Table SM.1** Values of crystallographic unit cell angles calculated for both $Ag_2F_5$ polymorphs, from DFT+U ($U_{Ag}$ = 5, 6 or 8 eV), SCAN and HSE06 methods.

|       | Angle | hse06    | scan     | U = 5 eV | U = 6 eV | U = 8 eV |
|-------|-------|----------|----------|----------|----------|----------|
| $C2/c$ | β     | 115.1572 | 115.2962 | 115.3741 | 114.9179 | 114.8356 |
|       | α     | 89.5719  | 89.352   | 89.3078  | 89.3484  | 89.5422  |
| $P\bar{1}$ | β  | 107.1009 | 105.5684 | 106.9928 | 107.0419 | 107.1179 |
|       | γ     | 90.7511  | 90.8127  | 90.8848  | 90.9042  | 90.7197  |

### III. Relative stability of both polymorphs at elevated pressure

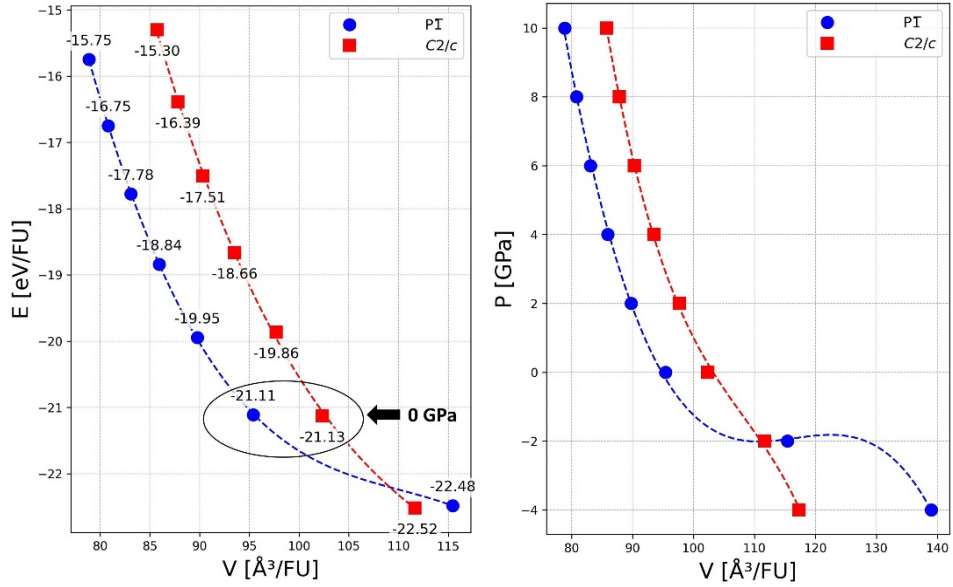

**Figure SM.2** Equation of state for both forms of $Ag_2F_5$ at 0 K (DFT+U, $U_{Ag}$ = 5 eV). On the left side of the figure, the static dependence of the energy per formula unit (FU) on volume is shown (dashed curves) for both structural forms. The circled minima correspond to zero external pressure. On the right side, the dependence of pressure (*p*) on volume (*V*) is shown. Red color for monoclinic *C2/c*, blue for $P\bar{1}$.

### IV. Magnetic interactions

In the DFT+U calculations, spin states analogous to those adopted for determining the magnetic coupling constants in $AgCoF_5$ were assumed[2].
The Hamiltonian employed for the calculation of the superexchange constants is presented below:

$$H = -2 \cdot J^b_{1(A-B)} \cdot (S^{A1} \cdot S^{B1} + S^{A3} \cdot S^{B3} + S^{A2} \cdot S^{B2} + S^{A4} \cdot S^{B4})$$
$$-2 \cdot J^a_{2(A-B)} \cdot (S^{A1} \cdot S^{B3} + S^{A3} \cdot S^{B1} + S^{A2} \cdot S^{B4} + S^{A4} \cdot S^{B2})$$
$$-2 \cdot J^c_{3(B-B)} \cdot (S^{B1} \cdot S^{B2} + S^{B3} \cdot S^{B4})$$
$$-2 \cdot J^c_{4(A-A)} \cdot (S^{A1} \cdot S^{A2} + S^{A3} \cdot S^{A4})$$
$$-2 \cdot J_{5(A-B)} \cdot (S^{A1} \cdot S^{B4} + S^{A3} \cdot S^{B2}) \quad \text{(eq. SM1)}$$

Where A = $Ag^{2+}$, B = $Ag^{3+}$, $S^A$ = ½, $S^B$ = 1

The energy values of the various spin configurations obtained from the calculations allowed for the determination of the magnetic coupling constants using the determinant method to solve the corresponding system of linear equations. The arrangements of spins in respective magnetic states, described as e.g. FM, AFM2 or AFM/FM, was analogous to those adapted in isostructural $AgCoF_5$[2]. The individual coupling constants were calculated according to the equations presented below:

$$J^b_{1(A-B)} = \frac{1}{8}(E_{AFM/FM5} - E_{FM}) \quad \text{(eq. SM2)}$$

$$J^a_{2(A-B)} = -\frac{1}{16}(-E_{FM} + E_{AFM1} - E_{AFM2} - E_{AFM/FM} + 2E_{AFM/FM5}) \quad \text{(eq. SM3)}$$

$$J^c_{3(B-B)} = -\frac{1}{8}(-E_{AFM/FM4} + \frac{1}{2}E_{FM} + \frac{1}{2}E_{AFM/FM}) \quad \text{(eq. SM4)}$$

$$J^c_{4(A-A)} = -\frac{1}{4}(E_{FM} - E_{AFM/FM2} + E_{AFM/FM}) \quad \text{(eq. SM5)}$$

$$J_{5(A-B)} = -\frac{1}{16}(-E_{AFM/FM} + E_{FM} - E_{AFM1} + E_{AFM2}) \quad \text{(eq. SM6)}$$

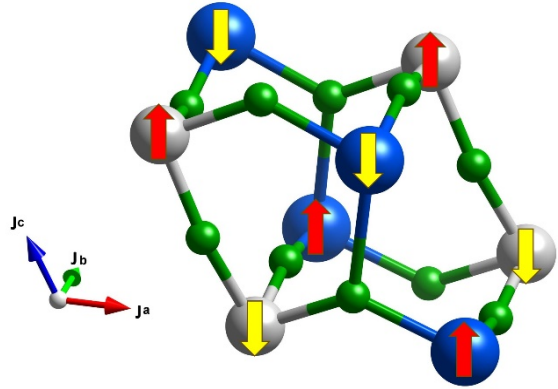

Figure SM3 G-type antiferromagnet, magnetic ground state of $Ag_2F_5$.

## V. List of frequency of vibrations at Gamma point for $P\bar{1}$ polymorph (DFT+U, $U_{Ag}$ = 5 eV)

| # | Frequency [cm$^{-1}$] | # | Frequency [cm$^{-1}$] | # | Frequency [cm$^{-1}$] | # | Frequency [cm$^{-1}$] |
|---|---|---|---|---|---|---|---|
| 1 | 584 | 22 | 336 | 43 | 198 | 64 | 109 |
| 2 | 580 | 23 | 334 | 44 | 196 | 65 | 104 |
| 3 | 575 | 24 | 300 | 45 | 191 | 66 | 95 |
| 4 | 564 | 25 | 296 | 46 | 186 | 67 | 94 |
| 5 | 556 | 26 | 296 | 47 | 178 | 68 | 93 |
| 6 | 545 | 27 | 281 | 48 | 175 | 69 | 92 |
| 7 | 544 | 28 | 275 | 49 | 172 | 70 | 87 |
| 8 | 519 | 29 | 271 | 50 | 162 | 71 | 83 |
| 9 | 511 | 30 | 266 | 51 | 160 | 72 | 78 |
| 10 | 503 | 31 | 265 | 52 | 152 | 73 | 75 |
| 11 | 496 | 32 | 261 | 53 | 149 | 74 | 70 |
| 12 | 495 | 33 | 258 | 54 | 145 | 75 | 67 |
| 13 | 491 | 34 | 242 | 55 | 141 | 76 | 63 |
| 14 | 482 | 35 | 240 | 56 | 139 | 77 | 60 |
| 15 | 474 | 36 | 238 | 57 | 132 | 78 | 57 |
| 16 | 471 | 37 | 232 | 58 | 130 | 79 | 45 |
| 17 | 443 | 38 | 231 | 59 | 126 | 80 | 33 |
| 18 | 437 | 39 | 225 | 60 | 118 | 81 | 29 |
| 19 | 433 | 40 | 216 | 61 | 115 | 82 | -1 |
| 20 | 402 | 41 | 215 | 62 | 114 | 83 | -1 |
| 21 | 337 | 42 | 202 | 63 | 112 | 84 | -2 |

## VI. Optimized structure of *C*2/*c* polymorph (DFT+U, $U_{Ag}$ = 5 eV)

Lattice parameters

| a | b | c | alpha | beta | gamma |
|---|---|---|---|---|---|
| 7.35180 | 7.86405 | 7.83626 | 90.0000 | 115.3741 | 90.0000 |

Unit-cell volume = 409.35 Å^3

|  | x | y | z |
|---|---|---|---|
| Ag1 | 0 | 0 | 0 |
| Ag2 | 0 | 0 | 0.5 |
| Ag3 | 0.5 | 0.5 | 0 |
| Ag4 | 0.5 | 0.5 | 0.5 |
| Ag1 | 0 | 0.5 | 0 |
| Ag2 | 0 | 0.5 | 0.5 |
| Ag3 | 0.5 | 0 | 0 |
| Ag4 | 0.5 | 0 | 0.5 |
| F1 | 0 | 0.94019 | 0.25 |
| F2 | 0 | 0.05981 | 0.75 |
| F3 | 0.5 | 0.44019 | 0.25 |
| F4 | 0.5 | 0.55981 | 0.75 |
| F5 | 0.80581 | 0.46357 | 0.62415 |
| F6 | 0.19419 | 0.53643 | 0.37585 |
| F7 | 0.19419 | 0.46357 | 0.87585 |
| F8 | 0.80581 | 0.53643 | 0.12415 |
| F9 | 0.30581 | -0.03643 | 0.62415 |
| F10 | 0.69419 | 0.03643 | 0.37585 |
| F11 | 0.69419 | -0.03643 | 0.87585 |
| F12 | 0.30581 | 0.03643 | 0.12415 |
| F13 | 0.02886 | 0.25164 | 0.45904 |
| F14 | -0.02886 | 0.74836 | 0.54096 |
| F15 | -0.02886 | 0.25164 | 0.04096 |
| F16 | 0.02886 | 0.74836 | 0.95904 |
| F17 | 0.52886 | 0.75164 | 0.45904 |
| F18 | 0.47114 | 0.24836 | 0.54096 |
| F19 | 0.47114 | 0.75164 | 0.04096 |
| F20 | 0.52886 | 0.24836 | 0.95904 |